# Design and Implementation of Curriculum System Based on Knowledge Graph

Xiaobing Yu[1]
*Miami University*
Oxford, USA
yux12@miamioh.edu

Han Chen[2]
*Capital University of Economics and Business*
Beijing, China
mhanchen@126.com

Mike Stahr[1]
*Miami University*
Oxford, USA
stahrm@miamioh.edu

Runming Yan[2]
*Miami University*
Oxford, USA
yanr5@miamioh.edu

*Abstract*—With the fact that the knowledge in each field in university is keeping increasing, the number of university courses is becoming larger, and the content and curriculum system is becoming much more complicated than it used to be, which bring many inconveniences to the course arrangement and analysis. In this paper, we aim to construct a method to visualize all courses based on Google Knowledge Graph. By analysing the properties of the courses and their preceding requirements, we want to extract the relationship between the precursors and the successors, so as to build the knowledge graph of the curriculum system. Using the graph database Neo4j [7] as the core aspect for data storage and display for our new curriculum system will be our approach to implement our knowledge graph. Based on this graph, the venation relationship between courses can be clearly analysed, and some difficult information can be obtained, which can help to combine the outline of courses and the need to quickly query the venation information of courses.

## I. INTRODUCTION

In view of the continuous development and improvement of the modern university curriculum system, the relationship between different courses under different professional categories becomes closer. Therefore, the design of knowledge graph of the curriculum system will effectively promote the research and professional development of the curriculum system, and facilitate teachers and students to consult and understand related courses and their main skills. Unlike the resource systems in other fields, the design of the curriculum system must take into account the types of courses, the relationship between courses and other major aspects who plays important roles in the system. It will be liable to deal with these tasks by manually sorting out or using SQL statements to achieve statistics, analysis and prediction of data. Knowledge graph is an effective way to solve these problems. The construction of knowledge map of curriculum system is of great significance to solve the following problems.

### A. Applying visual graphical interface to effectively detect the typing and scheduling errors in the course syllabus

The compilation of the course syllabus is often completed manually. In this process, there may be a variety of errors due to the negligence of the compilers and entrants. The application of knowledge graph can resolve the errors in the course syllabus, and provide effective support for the following construction of knowledge graph for the curriculum system and subsequent analysis work.

### B. Ability to effectively improve the efficiency of searching courses and combing their venous relationships

The construction of the course system involves a large number of courses and the context between them and the knowledge graph can effectively combine this information together, and transform the confused information sources into orderly and easy-to-use knowledge sources. By describing the semantic relationship in the graph, the scattered information in the course syllabus can be integrated and managed, and useful information and knowledge are discovered, selected and organized, and then transferred to the people or systems in need, so that the effective analysis of information can be achieved.

In this paper, we proposed a construction scheme of curriculum system based on knowledge graph. The design and implementation of knowledge graph for curriculum system are based on the Computer Science Department syllabus of Miami University [1]. We will start from the prerequisite courses to construct knowledge graph of curriculum system which will be systematically described, and used to detected and mined further issues based on the basis of experimental data on our target dataset. This paper also explores the existing problems in the curriculum syllabus, combs and analyses its context, queries and displays the precursors and successors of specific course nodes, and improves the efficiency of using the knowledge graph of the curriculum system.

## II. RELATED WORK

The concept of knowledge graph was first proposed by Google [2], which is essentially a structured semantic knowledge base that can be easily understood by users to graphically describe entities and their relationships in the real world, and forming a network of knowledge structures with the existing data and facts. Knowledge graph provides a practical and important reference for revealing the law of dynamic development in the field of knowledge, describing the logical relationship between each entity, and predicting the future development trend of the field, and this graph comes with the advantage of being more intuitive and easy to infer. It has been applied in many fields such as sports, movies and TV industry [3, 4]. However, its application in the field of curriculum system construction is rare.

We have searched some papers who have conducted with some research which shares similar thought with ours. One paper analyses the relationship between resource entities through modular semantic analysis and keyword extraction of digital course resources on the network, and finally draws a knowledge graph in this field [5], which implements a recommendation tool based on this specific knowledge graph, and we believe this study is very good at recommending related learning resources for novel learners. We believe this paper provides a new way of thinking for educators who study the knowledge graph for university curriculum system, and we will follow this thought and expand this idea with the actual implementation of knowledge graph [6].

This paper will put forward a construction scheme of the curriculum system by clearly describing the logical relationship between the courses based on the knowledge graph, the potential errors in the curriculum system can be found more effectively, and the arrangement errors can be helpful for users to sort out the context of the relationship between the courses. Users do not need to search a large number of related knowledge points, but view the relevant information of the course through the specified nodes in the knowledge graph, eliminating the tedious search steps, greatly improving the search efficiency, and providing convenience for modern instructing.

## III. METHODOLOGY

In order to make our knowledge graph reasonable and user-friendly, we need first organize the whole relationship between each course. We decided to utilize the course list from Computer Science and Engineer Department in Miami University, Oxford [1]. Our construction is summarized as follows:

### A. Constructing Knowledge Graph Model of Curriculum System

According to the name and relationship of the course, building the knowledge graph of the curriculum system requires attributes and related information of the course be obtained from structured data. By acquiring and combing this information, the type and entity are defined. This will expand the expression, storage and query of the knowledge graph of the curriculum system, as well as the error correction problems based on the knowledge graph.

### B. Acquiring Multi-Source Entity Knowledge

The data for this project to create knowledge graph are mainly from the syllabus of courses for multiple courses, and the professional training program for a given specialty [8]. Most of these data are well-organized since we retrieve all data from structured datasets. This paper will study how to extract these data precisely and accurately, then analyze the information contained in them to achieve the acquisition of multivariate entity knowledge.

### C. Entity Relationship

To create a knowledge graph of the curriculum system, the attributes of the course itself and the various relationships between the course and the course need to be considered. Through the analysis and summary of the course syllabus, these aspects can be divided into the following forms:

1. Course Title/ Category: The basic and most critical properties of the course, such as "Object-Oriented Programming".

2. Course Description: The content of the course, which is summarizing what knowledge this course is design to instruct.

3. Course Meeting Days and Time: When this course will be held.

4. Main Skill: Core skill which is used in the course, such as "Java".

5. Course Credits: Credits for the course.

6. Capacity and Enrollment: Capacity of this course and the number of students who are enrolled in this course.

7. Instructor: Professor who will teach this course.

8. Prerequisites: Prerequisites for the course, such as "Fundamentals-Programming & Problem Solving".

We are attempting to efficiently deal with the analysis and arrangement tasks of courses by combing the context of the curriculum system with the knowledge map. The prerequisite relationship between courses is an important content to obtain from the system, and it is also the most complicated external relationship between courses. We will use the dataset from CSE department of Miami University[1] which we mentioned previously, and extract the related courses and their relationships, to form the correlation between courses well-organized and establish specific knowledge graph of the curriculum system. The process of creating a knowledge graph is shown in Figure 1.

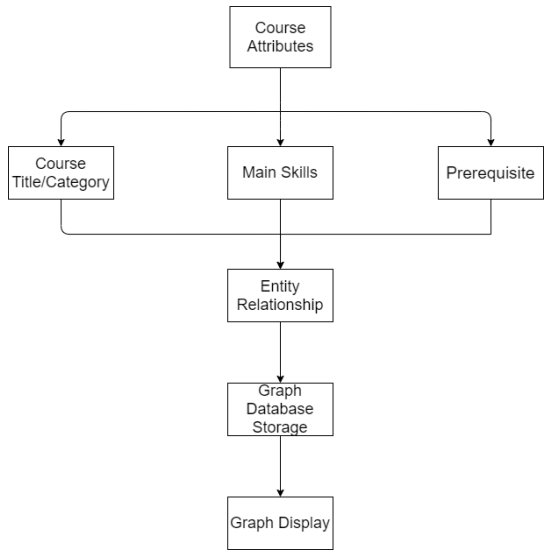

Fig. 1. The process of creating this knowledge graph

### D. Mapping of Knowledge Graph

*1) Data Preprocessing:* The context of the curriculum system is mainly composed of two parts, the curriculum syllabus and professional training program of each course, in which the curriculum syllabus often exists in the form of a WORD document, and the data in the WORD document includes the tabular form of the course attributes and the prerequisite relationship between the courses, while the data in the Excel document includes the training programs for CSE major and the distribution of course credits, both of which are structured data storage formats. Therefore, relevant content extraction rules can be formulated based on this feature. We take the course titles and the prerequisites as the final entity content to be extracted.

*2) Data Features Analysis:* The content of the course syllabus mainly includes the course title, the course category, the main skills, the prerequisites and other stuff. This paper selects the prerequisite as the main feature to extract the relationship between the course entities.

*3) Mapping of Knowledge Graph:* After the preparation from previous steps, we are capable to determine the relationship between all entities in the curriculum system, and then the data needs to be mapped as appropriate knowledge graph using appropriate tools. The graph database for knowledge graph includes Ucinet and SPSS. In this paper, we decided to take Neo4j, a popular graphics database, to complete the knowledge map for curriculum system [7]. When we use Neo4j to implement the knowledge graph, the first step is to store the data in the Neo4j graph database, where the data nodes are divided into different colors according to the data type in order to find the relationship between the nodes.

Here is a paradigm which shows a fragment of the knowledge graph of the curriculum system in Figure 2, in which the nodes such as "Intro to Software Engineering" and "Systems I" represent the corresponding course entities, and the relationship between courses can be visualized from the attributes in the diagram. For example, "Data Abstractions & Structures" is a prerequisite for "Algorithms I", but it is also a prerequisite for "Database Systems", "Comparative Program Languages" and other courses. The rest parts of the graph are similar.

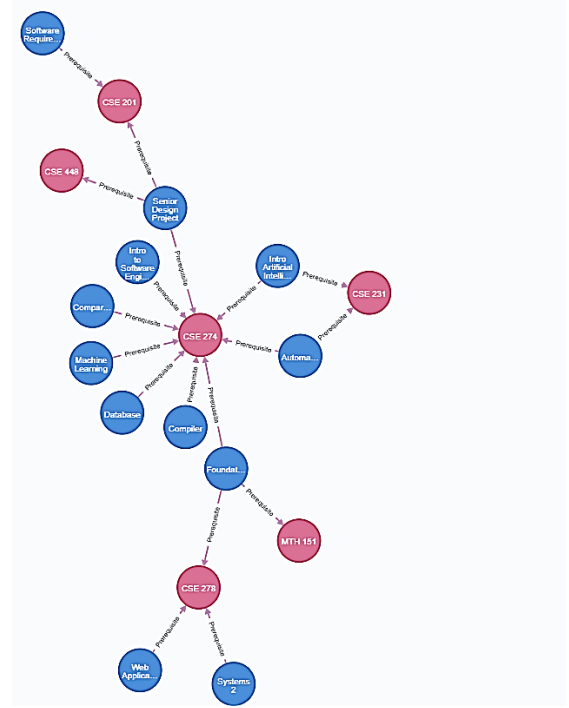

Fig. 2. Knowledge Graph of Part of the Curriculum System

## IV. DATA ANALYSIS BASED ON KNOWLEDGE GRAPH OF CURRICULUM SYSTEM

We design this knowledge graph according to the previous definitions and rules based on the experimental data of the enrollment and training program of major categories and the curriculum syllabus of 31 courses in Computer Science & Engineering Department of Miami University in Oxford [1]. We summarize this knowledge graph into following main functions.

### A. Query Function

Querying is the key function of the knowledge graph of the curriculum system, and it is also the key to ensuring that users can successfully find and sort out the course context. We can obtain a standard and complete graph database of the curriculum system by creating the knowledge graph which follows the previous stages. We already have realized the query function of the curriculum based on the existing database.

This function is designed to help users quickly find courses and their precursor-successor relationships. Queries can be divided into querying the precursor-successor of a single course and querying the venation relationship between two courses. After a user enters a specified course, the system normally will be able to match in the knowledge graph of the curriculum system to find the precursor course of that course or the course that is the precursor course. When the user enters two courses; the system can find out the direct relationship between the two courses.

## B. Error Correction Function

In the process of data import, users will be alerted to errors in the course syllabus, such as typo on the title of a course, absence of the prerequisite, or other problems related to document writing and course veneering. The errors will be listed before they are checked and corrected by the user to continue with the import operation to generate a new knowledge graph of the curriculum system.

## C. Visualization

Visualization is to display the results of the current query, including the course title, the prerequisite, the course category and other information. We want to utilize this information to detect the deep relationship between courses, and apply our observation on further data analysis, which has important reference value for users of the knowledge graph. Figures 3 and 4 show the context between the prerequisites for a given course and the two specified courses respectively.

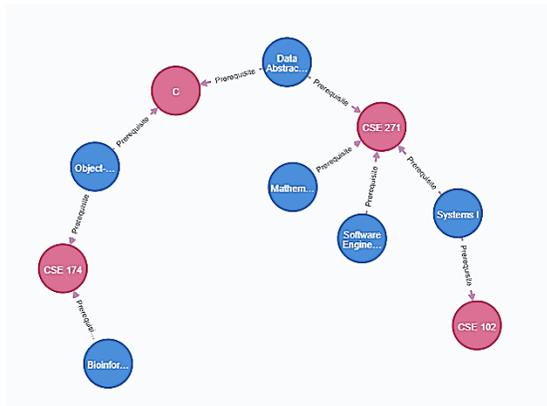

Fig. 3. The context between the prerequisites for a given course

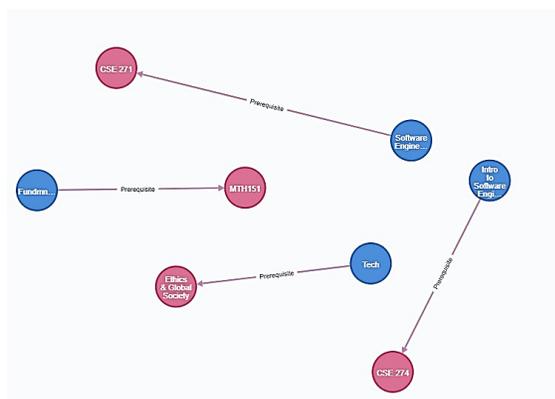

Fig. 4. The context between two specified courses

## V. CONCLUSION

In view of the problems such as insufficient intuitive display of the curriculum system, insufficient clarity of the context, time-consuming and laborious manual inspection of errors in the implementation of the curriculum syllabus, we propose to construct a knowledge graph based on the curriculum system by extracting the precursors and successors of the curriculum outline, so that students and teachers can organize the context and error information of the curriculum system. The knowledge and mastery of this integration will greatly help its overall data analysis and error checking, and provide a new way of thinking for the future arrangement and adjustment of the curriculum system.

The tasks are done in this paper will be parts of the research work we are doing, and they only deal with the relationship between the preceding and following parts of the course because the amount of data is not very large. The construction of this knowledge graph is specified with the curriculum content from CSE department of Miami University [1]. In the future, the other contents of the course will be taken into account and a larger scale will be built. We will expand our knowledge graph to enable more comprehensive association analysis and relationship mining of the curriculum system and to apply it to a broader area.